\documentclass[useAMS,usenatbib]{mn2e}

\usepackage{epsfig}
\usepackage{myaasmacros}

\title[Flow-Driven Cloud Formation in Eulerian and Lagrangian Simulations]
{Flow-Driven Cloud Formation and Fragmentation: Results From
Eulerian and Lagrangian Simulations}
\author[Heitsch et al.]{Fabian Heitsch$^{1,2,5}$\thanks{E-mail:
fheitsch@unc.edu;
naab@mpa-garching.mpg.de;
stefanie.walch@astro.cf.ac.uk},
Thorsten Naab$^{3,5}$, Stefanie Walch$^{4,5}$ \\
$^{1}$ Department of Physics \& Astronomy, UNC Chapel Hill, 120 E Cameron St, Chapel Hill,
NC 27599-3255, USA\\
$^{2}$ Department of Astronomy, University of Michigan, 500 Church St,
Ann Arbor, MI 48109-1042, USA\\
$^{3}$Max Planck Institut f\"ur Astrophysik, Karl-Schwarzschild-Str. 1, 85741 Garching, Germany\\
$^{4}$School of Physics \& Astronomy, Cardiff University, 5 The Parade, Cardiff CF24 3AA, UK\\
$^{5}$Universit\"ats Sternwarte M\"unchen, Scheinerstr.1, D-81679 M\"unchen, Germany}

\begin{document}

\date{Accepted March 11, 2011}

\pagerange{\pageref{firstpage}--\pageref{lastpage}} \pubyear{2010}

\maketitle

\label{firstpage}

\begin{abstract}
The fragmentation of shocked flows in a thermally bistable medium 
provides a natural mechanism to form
turbulent cold clouds as precursors to molecular clouds. Yet because of the 
large density and temperature differences and the range of
dynamical scales involved, following this process with numerical
simulations is challenging. We compare two-dimensional simulations 
of flow-driven cloud formation without self-gravity, using the Lagrangian Smoothed Particle
Hydrodynamics (SPH) code VINE and the Eulerian grid code Proteus. 
Results are qualitatively similar for both methods, yet the variable spatial resolution of 
the SPH method leads to smaller fragments and thinner filaments, rendering the overall
morphologies different. Thermal and hydro-dynamical instabilities lead to rapid cooling and 
fragmentation into cold clumps with
temperatures below $300$~K. For clumps more massive than $1$~M$_\odot$~pc$^{-1}$, the 
clump mass function has an average slope of $-0.8$.
The internal velocity dispersion of the clumps is nearly an order of magnitude
smaller than their relative motion, rendering it subsonic with respect to the
internal sound speed of the clumps, but supersonic as seen by an external observer.
For the SPH simulations most of the cold gas resides at temperatures
below $100$~K, while the grid-based models show an additional, substantial component
between $100$ and $300$~K. Independently of the numerical method our models confirm
that converging flows of warm neutral gas fragment rapidly and form high-density, 
low-temperature clumps as possible seeds for star formation.
\end{abstract}


\section{Introduction}
The highly filamentary morphology of molecular clouds (MCs) and their
observed non-thermal line-widths
\citep{1990ApJ...359..344F,2000prpl.conf...97W} point to MCs being
highly dynamical objects. Observational evidence 
suggests that star formation in local MCs such as Taurus
is rapid once molecular gas is available, and that the 
parental clouds are short-lived
\citep{2000ApJ...530..277E,2001ApJ...562..852H,2003ApJ...585..398H}. 
The dynamical, or turbulent, nature of MCs is assumed to play a
crucial role in the process of star formation via turbulent
fragmentation \citep[][see also \citealp{2006MNRAS.372..443B}, 
for a summary of the effects and interpretation of turbulence in MCs]{1981MNRAS.194..809L,2004RvMP...76..125M}.


Because of the rapid onset of star formation, the cloud {\em formation}
process needs to provide the MC with the observed turbulence and
substructure. Moreover, global geometry and gravity considerations
mandate that this substructure be non-linear \citep{2004ApJ...616..288B}, 
i.e. a physical process is needed that can imprint non-linear density perturbations
in the proto-cloud {\em during its formation}. These requirements have led
to the scenario of flow-driven cloud formation, where molecular clouds are assembled
by large-scale converging flows of atomic hydrogen corresponding to the 
warm neutral medium \citep[][see also 
\citealp{1993ApJ...419L..29E,2000ApJ...530..277E}]{1999ApJ...527..285B,2001ApJ...562..852H,2003ApJ...585..398H}.
The rapid fragmentation is driven by a combination of strong thermal and dynamical
instabilities, dominated by the thermal instability 
\citep[TI, ][see also \citealp{2008ApJ...683..786H} for a discussion of timescales]{1965ApJ...142..531F}.

Large-scale gas flows are ubiquitous in the Galaxy. They might be
driven locally by supernova explosions \citep{1977ApJ...214..725E,1987ApJ...317..190M,2008PASP..120..972N} 
or globally by shear motions in the Galactic disk, by global graviational instabilities \citep{2007ApJ...671..374Y}, 
gas infall from the halo \citep{1985ApJ...290..154L,1982ApJ...256..112M}, interactions 
with the central bar \citep{1979ApJ...233...67R,1985A&A...150..327C} or satellite
galaxies. On extragalactic scales collisions of galactic disks trigger gas flows, shocks and starbursts
\citep{1996ApJ...464..641M,2006MNRAS.372..839N,2010ApJ...715L..88K}. 

High-resolution simulations in two
(\citealp{2005A&A...433....1A,2007A&A...465..445H,2007A&A...465..431H}; 
\citealp{2005ApJ...633L.113H,2006ApJ...648.1052H}) and three dimensions
(\citealp{2006ApJ...643..245V,2007ApJ...657..870V}; \citealp{2008ApJ...674..316H};
\citealp{2008A&A...486L..43H}; \citealp{2009MNRAS.398.1082B})
have demonstrated that the flow-driven formation of clouds is indeed
a natural and elegant way to provide the clouds with the observed
turbulence and substructure. Except for \citet{2007ApJ...657..870V},
the above authors used various grid-based methods. However, due to the
high density contrast and the strong TI in the intermediate temperature
regime between $300< T< 5000$~K, the spatial scales of the forming cold filaments
and clumps shrink dramatically. This 
problem becomes even more severe with gravity acting upon the cold regions.

Due to its Lagrangian nature, the strength of Smoothed Particle Hydrodynamics
\citep[SPH, e.g.][]{1992ARA&A..30..543M} resides in its capability to follow fluid flows
at ``arbitrary'' spatial resolution. In particular, dissipative properties of SPH methods do
not depend on geometry or direction, which can cause issues 
for grid-based methods without physical dissipation control 
\citep{1998A&A...332..969R}. SPH has been used to model the interstellar medium on all
scales, from planet formation \citep[e.g.][]{2002Sci...298.1756M}, 
formation of protostellar disks from rotating and turbulent cores \citep{2009MNRAS.400...13W,2010MNRAS.402.2253W},
formation of stars and cores \citep{1997MNRAS.292...11K,2003MNRAS.339..577B},
ionization feedback from massive stars \citep{2009MNRAS.393...21G,2010ApJ...723..971G,2009A&A...497..649B},
formation of MCs in galactic disks 
\citep{2006MNRAS.371.1663D,2007MNRAS.376.1747D,2008MNRAS.389.1097D}, galaxy evolution
\citep{1996MNRAS.278.1005S,1989ApJS...70..419H,2000MNRAS.312..859S,2006MNRAS.372..839N,2007ApJ...658..710N},
to large scale simulation of galaxy formation  
\citep[e.g.][]{2003MNRAS.339..289S}. \citet{2007ApJ...657..870V} present models of 
flow-driven MC formation using SPH. However, SPH has its own
inherent limitations, among others a limited mass resolution, and
spatially varying dissipative properties. Hence the question remains
whether SPH and grid-based methods give similar results for a specific
astrophysical problem, and what the method-intrinsic differences are.  
\citet{2005ApJ...627..608N} showed for cosmological applications that
both approaches give statistically similar results, and
\citet{2000ApJ...535..887K} concluded the same for models
of self-gravitating driven turbulence in a periodic box of isothermal gas,
although \citet{2009A&A...508..541K} show for the latter application that SPH is 
more dissipative than grid models for decaying turbulence.

Yet it remains unclear whether this agreement between methods holds for models of a thermally
unstable, turbulent interstellar medium. This is of particular interest as 
\citet{2007MNRAS.380..963A} and \citet{2010MNRAS.407.1933J} have pointed out SPH's limitations
evolving shear flow instabilities \citep[see, however, ][]{2008JCoPh.22710040P,2010MNRAS.405.1513R,2010arXiv1003.0937A} -- a
crucial ingredient in the formation of turbulent clouds from colliding
flows \citep{2006ApJ...648.1052H,2008ApJ...683..786H}.

This paper compares the Eulerian and Lagrangian approach to the problem
of flow-driven cloud formation in two dimensions. The concept 
of the simulations follows \citet{2005ApJ...633L.113H,2006ApJ...648.1052H} and is summarized in
\S\ref{MODELS}. We find (\S\ref{RESULTS}) a good agreement between SPH
and grid-based results in terms of the rapid onset of fragmentation and, to a lesser
degree, of morphology. 

The total amount of mass accumulated in the cold gas component ($T < 300$~K) is smaller in 
the SPH models, yet the temperature distribution within the cold gas is concentrated towards 
lower temperatures. The grid models show a wider spread of temperatures, but an overall higher 
mass in cold gas. We attribute this difference to the over-cooling problem in SPH
described by e.g. \citet{1999ApJ...521L..99P}. Line-of-sight velocity
dispersions as seen by an observer are generally supersonic with respect to the cold gas,
yet, the dispersions within the cores are approximately sonic, in agreement with observations
\citep[e.g.][]{2007A&A...472..519A}.
The strength of the Lagrangian approach is demonstrated when
deriving the mass distribution of cold dense fragments forming in the
colliding flows.

\section{The models}
\label{MODELS}


All SPH simulations are run with the parallel shared memory SPH code 
VINE \citep{2009ApJS..184..326N,2009ApJS..184..298W}. 
The code solves the asymmetric form 
of the energy equation and uses the standard implementation of the 
artificial viscosity to treat shocks
\citep{1992ARA&A..30..543M}. SPH particles have individual
time-steps that can cover approximately eight orders of magnitude. 
The time steps are determined by the local Courant-Friedrichs-Lewy
criterion \citep{monaghan89} ensuring that information is not
carried over length scales larger than the local kernel size that is
defined by the 20 nearest neighbor particles in two dimensions.  
In addition we require that the fractional change of the kernel size
of each particle is limited to 10 percent. Thereby we ensure that
particles require several time steps to pass through the interface of
a strong shock. 

For the fixed-grid simulations we used the gas-kinetic grid code
Proteus \citep{1993JCoPh.109...53P,1999A&AS..139..199S,2001JCoPh.171..289X,2004ApJ...603..165H}
in an identical implementation as described by \citet{2008ApJ...674..316H}. Proteus allows
the explicit control of viscosity and heat conductivity at fixed spatial scales
\citep[see][for a detailed description]{2002MNRAS.333..894S}.

\subsection{Heating and Cooling}\label{ss:heatcool}

To compute the heating and cooling rates we have used the rates for
optically thin atomic lines from \citet{1995ApJ...443..152W,2003ApJ...587..278W}. The
cooling curve covers a range of $10^{-2} \le n \le 10^3$~cm$^{-3}$
in density and $30\mbox{K} \le T \le 2.5 \times 10^4$~K in temperature and is 
identical to that used by \citet{2006ApJ...648.1052H}.
Dust extinction becomes important above
column densities of $N(\mbox{HI})\approx 1.2\times 10^{21}$cm$^{-2}$, which are
reached only in the densest regions modeled. Thus, we use the unattenuated
UV radiation field for grain heating,
expecting substantial uncertainties in cooling rates only for the densest regions.
The ionization degree is derived from a balance between ionization by cosmic rays and
recombination, assuming that Ly $\alpha$ photons are directly reabsorbed.
As in \citet{2006ApJ...648.1052H}, heating and cooling is implemented iteratively as a source
term for the internal energy $e$ of the form
\begin{equation}
  \frac{de}{dt} = n\Gamma(T) - n^2\Lambda(T)
  \label{e:cooling}
\end{equation}
in units of energy per volume per time.
Here, $\Gamma$ is the heating contribution (mainly photo-electric heating from grains),
and $n\Lambda$ is the cooling contribution (mainly due to the CII line at $158\mu$m).
To speed up the calculations, equation~(\ref{e:cooling}) is tabulated on a $2048^2$ grid
in density and temperature. For each cell and iteration, the actual energy change
is then bilinearly interpolated from this grid.

The implementation of heating and cooling in VINE follows closely that for Proteus 
described above. 
For each particle $i$ the change of the specific internal energy, $u$,
in units of energy per unit mass per unit time is computed as 
\begin{eqnarray}
n_i\frac{du_i}{dt}= n_i\Gamma(T_i) - n_i^2\Lambda(T_i) \label{cooling}
\end{eqnarray}
If the current integration time-step of the particle is larger than one percent
of the local cooling time the cooling is computed iteratively in
steps of one percent of the local cooling time. Using periodic boxes
of constant density and temperature at different values we have tested
that this implementation results in the correct cooling rates. 
 
\subsection{Initial and Boundary conditions}\label{ss:boundcond}
Two opposing, uniform, identical flows in the $x$-$y$ computational
plane initially collide head-on at a sinusoidal interface with 
wave number $k_y=1$ and amplitude $\Delta$ \citep[see Fig.~1 in][]{2006ApJ...648.1052H}.
The incoming flows are in thermal
equilibrium at a density of
$n_0=1.0$~cm$^{-3}$, a temperature of
$T_0=8.5\times10^3$~K and a velocity of $21$ km~s$^{-1}$, typical
for parameters of the warm neutral medium. The box length is $44$~pc.

We ran identical model sets with VINE (model names headed by ``V'') and Proteus
(names headed by ``P''). The model name V0256 specifies
an SPH model with initially $256^2$ particles, while e.g. P2048 denotes a
grid model with $2048^2$ cells.

Initially we distribute the SPH particles on a grid in the square domain.
The particle mass resolution is $7.2\times 10^{-4}$~M$_\odot$~pc$^{-1}$.
We create new particle layers at the $x$-boundaries at the initial
grid resolution with the appropriate velocities. Thus, the total particle
number of the simulation is continuously increasing, starting with 
with $256^2= 65536$ particles. At $9$ Myr, there are already
$640,000$ particles in model V0256.
This leads to a significant slowing down as the
simulation proceeds. 

We use periodic boundary conditions in the
y-direction. 
\citet{2006ApJ...648.1052H} explored the effect of 
the choice of boundary conditions on the mass budget and the morphology
of the forming clouds. They found that although open boundaries allow
a small fraction of the gas to leave the simulation domain (or active
region), the total mass budget and the morphology is only slightly affected.
The cooling timescale is much shorter than the flow timescale and the 
sound crossing time, so that the gas is compressed and cooled down before
it can feel the boundaries. Only gas close to the boundaries will be affected.

\section{Results}
\label{RESULTS}

\subsection{Morphologies}\label{ss:morphologies}

\begin{figure*}
\begin{center}
  \epsfig{file=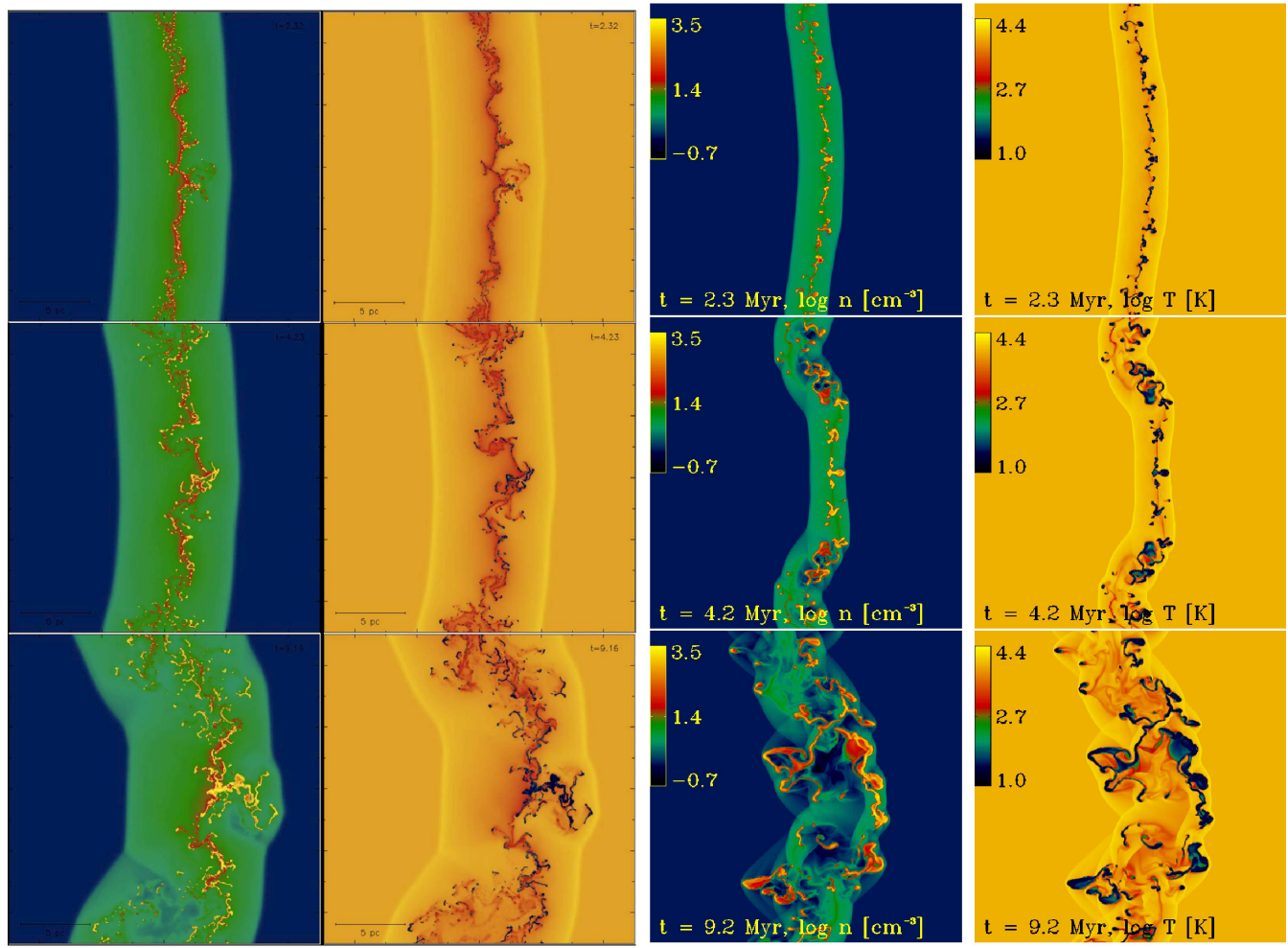,width=\textwidth}
  \caption{Morphologies of colliding flows. Shown are logarithmic density and
           temperature maps of the central quarter of the
           simulation domain, at times $t=2.3, 4.2$ and $9.2$~Myr, {\em left} for
           model V0256, and {\em right} for model P1024. The effective spatial resolution
           of V0256 leads to more filigree structure than in model P1024, with fixed resolution. 
           Densities are scaled identically between $-0.7 < \log n < 3.5$ for both models, and temperatures
           between $1.0 < \log T < 4.4$.
           \label{temp_dens_snap}}
\end{center}
\end{figure*}

We begin with the flow morphologies (Fig.~\ref{temp_dens_snap}), comparing 
density and temperature maps of models V0256 (left) and P1024 (right).
Densities and temperatures have been scaled to the same ranges, and for both codes, 
the central quarter ($22^2$~pc$^2$) of the simulation domain is shown at times 
$t=2.3,4.2$ and $9.2$ Myr after initial flow contact.  
The most noticeable difference is the much more filigree structure of the cold
gas exhibited in the SPH simulations. 
Comparing the initial resolution ($256^2$ vs $1024^2$) against the 
effective resolution indicates the resolving power of the Lagrangian method.
The smallest smoothing lengths in model V0256 correspond to $2\times 10^{-3}$
pc, while the grid resolution of model P1024 is $4\times 10^{-2}$~pc, and that
of P8192 $5.3\times10^{-3}$~pc. 

The SPH models tend to show a broader post-shock region (greenish colors) than 
the grid models. 
The densities for V0256 and P1024 are scaled
identically (between $-0.7 < \log n < 3.5$), as well as the temperatures 
(between $1.0 < \log T < 4.4$). V0256 clearly shows thinner filaments, but at approximately
the same density range, as model P1024 -- a consequence of the density-dependent spatial
resolution of SPH.

\begin{figure}
  \epsfig{file=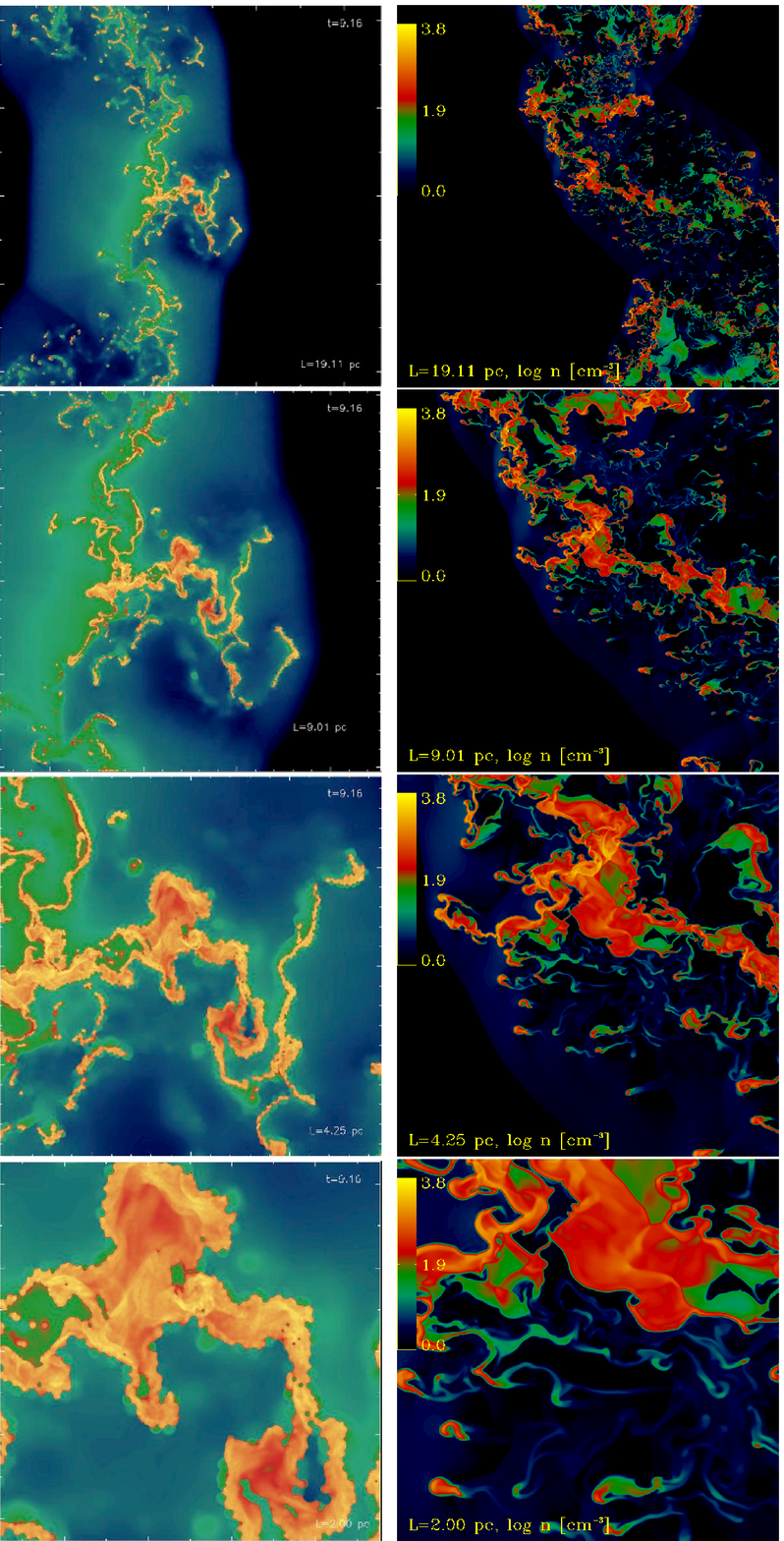,width=\columnwidth}
  \caption{Zoom sequence of V0256 (left) and P8192 (right), logarithm of density, shown at $t=9.2$~Myr,
           at (from top to bottom) a scale of $19$, $9$, $4.25$, and $2$~pc. Color scales are identical
           for both models, highlighting the differences.
           The full simulation domain would measure $44$~pc a side. The smallest dense structures
           (at a zoom of $2.0$~pc) have a size of $<0.1$~pc. Turbulent wakes are
           clearly visible in model P8192.\label{zoomseq}}
\end{figure}

Figure~\ref{zoomseq} shows a zoom sequence of models V0256 (left) and P8192 (right). The spatial extent
of each zoom level is indicated in the lower left of the respective panel.
The spatial resolution of model P8192 ($5.3\times 10^{-3}$~pc)
is only a factor of $2$ away from resolving the Field length \citep{1965ApJ...142..531F} 
for a realistic value of the heat conductivity \citep{2004ApJ...602L..25K}.
A closer study of the corresponding temperature maps (not shown) indicates that
the cooling length especially between the warm ambient and cold dense gas
is resolved in P8192: the cold dense regions start to show ``halos'' of intermediate temperature
material.

Comparing V0256 and P8192 at the largest scale ($L=19$~pc), it is already obvious that the SPH model 
reaches higher densities, at a more filamentary appearance. Conversely, stuctures in P8192 seem to 
be more extended, and less filamentary. This changes when moving to smaller scales (down to $2$~pc).
V0256 exhibits mostly coherent filaments at high densities ($n>10^2$~cm$^{-3}$), while P8192 shows a
whole population of small cloudlets including turbulent wakes. These are conspicuously absent in 
V0256. Thus, while the density-dependent spatial resolution allows for higher density contrasts and smaller
structures in V0256, the constant spatial resolution in P8192 leads to a better tracking of small 
low-density structures. The differences suggest (compared to the density ranges in
Fig.~\ref{temp_dens_snap}) that VINE reaches an effective grid resolution in the dense gas of $1024^2$ for
(initially) $256^2$ particles. The highest zoom also suggests that the density contrasts are more sharply
resolved in P8192 than in V0256, leading to smoother gradients in the SPH model.

The highly structured ``slab'' of model P8192 and a comparison with P1024
shown in Figure~\ref{temp_dens_snap} emphasizes the importance of resolution
for the problem of turbulent thermal fragmentation: only at
high Reynolds numbers, a turbulent cascade can develop.
Proteus runs with a specified viscosity, measuring 
$\nu=8.9\times10^{22}$~cm$^2$~s$^{-1}$ for our parameters. Using half the 
period of the geometric perturbation in the collision interface as typical 
length scale, and the inflow velocity as typical flow velocity, results in 
a formal Reynolds number of $Re\approx 1.6\times10^3$ for P1024, and 
$Re\approx 1.3\times 10^4$ for P8192.
Since the condensation mode
of the TI responsible for the fragmentation grows
first on the smallest scales \citep{2000ApJ...537..270B}, the generation
of small-scale structures due to turbulence on which the TI can feed will
be crucial for the overall evolution of the system. 

\subsection{Mass Fractions}\label{ss:massfrac}

\begin{figure*}
\begin{center}
  \epsfig{file=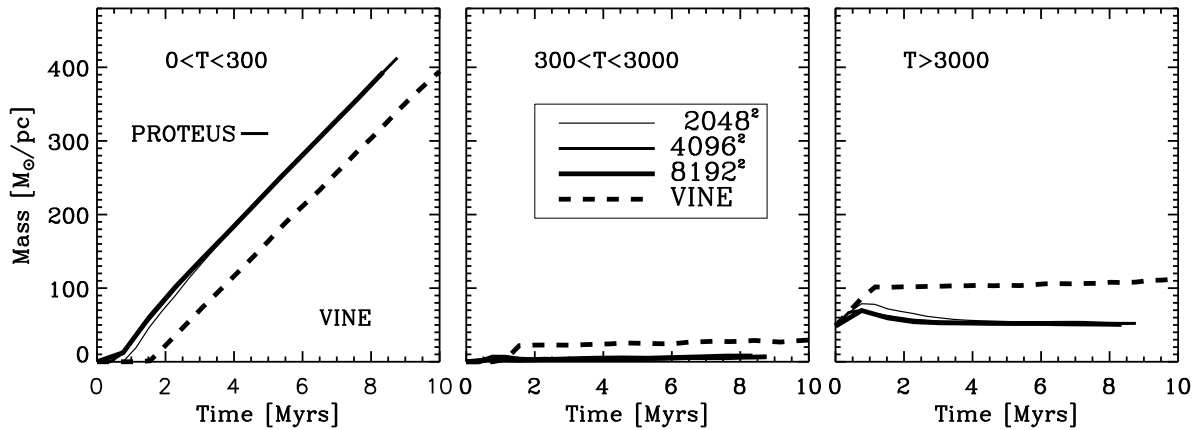,width=0.9\textwidth}
  \caption{Total mass per length in the three temperature regimes
           $T<300$~K, $300<T<3000$~K and $3000$~K $<T$ 
           as a function of time. Solid thick lines denote the grid models
           P2048, P4096 and P8192, while dashed line shows model V0256.
           Mass histories for models P4096 and P8192 are indistinguishable.
           The constand mass-offset between VINE and Proteus models is due to
           the higher artificial viscosity in VINE.
           \label{time_vs_massfrac}}
\end{center}
\end{figure*}

Figure~\ref{time_vs_massfrac} shows the mass per length for the
cold thermally stable regime ($T<300$~K), the warm thermally unstable
($300<T<3000$~K), and the warm thermally stable regime ($3000$~K$<T$),
as a function of time. 
For both codes, the mass of cold gas increases nearly linearly with time once
the thermal instability has set in.
Thus, despite all its substructure, the interaction zone acts like a one-dimensional
slab in terms of mass collection. The slight offset in time between the 
SPH and grid models is caused by the stronger effect of the artificial 
(bulk) viscosity in SPH to prevent the penetration of particle layers.
The gas thus reaches higher temperatures and requires more time to arrive
at densities beneficial for strong cooling (and mass accumulation). 
This leads to a constant mass offset of $\approx 70$~M$_{\odot}$~pc$^{-1}$
in the cold gas between Proteus and VINE models, corresponding to
$\approx 20$~M$_\odot$~pc$^{-1}$ less intermediate temperature gas and 
$\approx 50$~M$_\odot$~pc$^{-1}$ less high temperature gas in the Proteus
models. Mass histories of models P4096 and P8192 are indistinguishable.

While the SPH models show less mass in the cold temperature regime, there
is more mass in the intermediate and warm regime. Both codes evolve
the warm and intermediate regime qualitatively similarly: the mass in both
regimes stays approximately constant, since they are only transitory stages
of the gas on its way to the cold regime.

Considering the widely differing numerical methods, the mass fractions agree
surprisingly well. Especially the problem of over-cooling in SPH-methods
\citep{1999ApJ...521L..99P} does not seem to dominate the mass budget.

\subsection{Clump Mass Function}\label{ss:cmf}

The strong fragmentation due to the thermal instability begs the question what
the resulting clump mass function (ClMF)\footnote{We use the abbreviation ClMF to distinguish 
between clump and core mass function, the latter of which is usually abbreviated as CMF in the literature.} 
of the cold clouds will be. While the
detailed physical processes linking the ClMF to the core mass function, and
eventually to the initial mass function are still a matter of debate
\citep[e.g.][]{2000ApJS..128..287K,2000prpl.conf...97W,2002ApJ...576..870P,
2006ApJ...637..384B,2007ASPC..362..269E,2007A&A...465..431H,2008ApJ...678L.105D,2009MNRAS.396..830S},
a model of rapid star formation in colliding flows should explain what
physical processes are setting the ClMF during the
formation of the molecular cloud. The two main fragmentation agents
are cooling and self-gravity, the latter of which we have
not implemented in the current models. While we envisage self-gravity
to be important on global scales \citep{2004ApJ...616..288B,2007ApJ...654..988H}
for finite clouds (i.e. non-periodic boundaries), on local
scales, the gas has to cool substantially before the Jeans lengths get
small enough for gravity to dominate. Which mechanism then dominates
setting the ClMF? Because of the early and 
strong fragmentation due to thermal (and turbulent) effects,
the ClMF might well be determined early on in the cloud evolution
(see also \citealp{2007A&A...465..445H} for 2D models, and
\citealp{2008ApJ...674..316H} for 3D models).

\begin{figure}
\begin{center}
  \epsfig{file=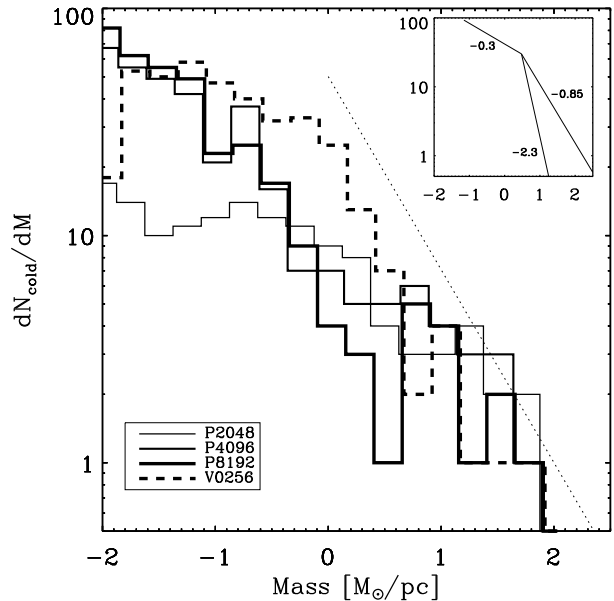,width=\columnwidth}
  \caption{\label{f:cmf}Histogram of clump masses (ClMF) for
           VINE (V0256,dashed line) and Proteus (P2048, P4096, P8192, solid lines), at $t=9.2$~Myr. The dotted line 
           shows a slope of $-0.85$ as determined for massive clumps in the Orion Molecular Cloud
           \citep{2007ApJ...655..351L}.} 
\end{center}
\end{figure}

The main strength of VINE over Proteus is of course the adaptive spatial resolution.
To identify clumps of cold gas in model V0256 we have used
a friends-of-friends algorithm with a linking length of $0.04$~pc, which
is about ten times the minimum kernel softening length at
this time, and a minimum of 20 particles per clump. In total we
found 418 individual clumps of gas with temperatures $T<300$~K.
The ClMF is plotted in
Fig. \ref{f:cmf}, at a model time of $9.2$~Myr. 
For comparison we have computed the ClMF
of model P2048, P4096 and P8192 using the same method. Instead of particle
positions, we used the cell positions of the cold gas as input for the
friends-of-friends algorithm, again with a linking length of $0.04$~pc.
For the Proteus models, the linking length is a factor $2$ to $8$ larger than
the (linear) cell size. As expected, the number of identified clumps increases for
the grid simulation with resolution ($153$ clumps for P2048, $495$ for P4096, and $860$ for P8192).
At high masses, P8192 and V0256 agree, whereas there is a clear deficit in clumps
at lower masses for the grid models, with a factor of three difference between V0256 and P8192 at 
$M\approx 1$~M$_\odot$~pc$^{-1}$.

A direct comparison of our ClMF to observations is problematic, for
two reasons. First, the restriction to two dimensions will emphasize compression (cooling) over
shear flows and vorticity (turbulence) since the gas flows are more compressible than in three dimensions 
\citep[see Fig.~13 of][]{2006ApJ...648.1052H}.
Second -- as the referee pointed out --, we chose to calculate ClMFs at a time ($9.2$~Myr) at which a sufficient
amount of gas will have accumulated for self-gravity to affect the evolution of the cloud. This can
be seen comparing models Hf1 and Gf1 of \citet[][see their Fig.~2]{2008ApJ...689..290H} at (their) $9.9$~Myr.
We chose this rather late time for our two-dimensional models, since the suppression of the third dimension
leads to rather coherent sheets filaments still imprinted by the initial conditions at earlier times.
 
Keeping the above caveats in mind, the SPH slope is consistent with
the clump mass slopes found by \citep[][see also Fig.~23 of \citealp{2009ApJ...698..324R}]{2007ApJ...655..351L} 
for Orion, and with earlier measurements by \citet{1998A&A...331L..65H} or \citet{2002A&A...384..225S}. 
Overall, the slopes of both methods are shallower than those observed for low-mass star forming regions such as 
Perseus \citep{2010MNRAS.402..603C}. Yet these authors point out that such differences might well be caused by the way in which clumps are defined,
since different clump-finding algorithms can result in very different slopes. In any case, for the reasons
given above,
our argument here focuses more on the comparison between the numerical methods (for which we use the identical
clump-finding mechanism), rather than on a detailed comparison with observations. 

\subsection{Line-of-sight Velocity Distributions}\label{ss:losvd}

\begin{figure}
\begin{center}
  \epsfig{file=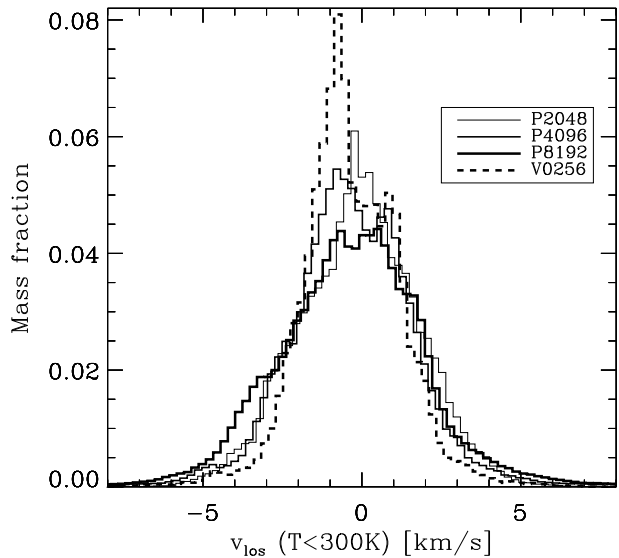,width=\columnwidth}
  \caption{Line-of-sight velocity distribution of the cold gas ($T<300$K)
  along the inflow direction. Dashed line: VINE, solid line: Proteus.\label{f:losvd}}
\end{center}
\end{figure}

In Fig \ref{f:losvd} we show the mass-weighted line-of-sight velocity
distribution of the cold gas ($T<300$~K) taken along the inflow direction. 
The velocity dispersion is $\approx 6.8$~km~s$^{-1}$ for VINE and 
$\approx 8.5$~km~s$^{-1}$ for Proteus. The sound speed of the cold gas
is $c_s\approx 0.75$~km~s$^{-1}$ at $T=40$~K, i.e. the cold gas motions are
nominally supersonic. 

While the overall agreement of the distributions
is reassuring, the differences are caused by resolution effects:
At higher resolution, smaller scales start to thermally fragment first 
(e.g. \citealp{2000ApJ...537..270B}), suppressing the growth of
the dynamical instabilities 
(e.g. \citealp{1997ApJ...483..262V}; \citealp{2003NewA....8..295H}). 
Since the SPH models have higher spatial resolution in their evolved state, we expect them
to be more dominated by thermal effects rather than by the 
dynamical instabilities, thus leading to a (slightly) smaller
velocity dispersion \citep[see also][]{2005ApJ...633L.113H}. Also, we expect 
the generally higher dissipative nature of SPH compared to grid methods 
\citep[e.g.][]{2009A&A...508..541K} to contribute.

\begin{figure}
\begin{center}
  \epsfig{file=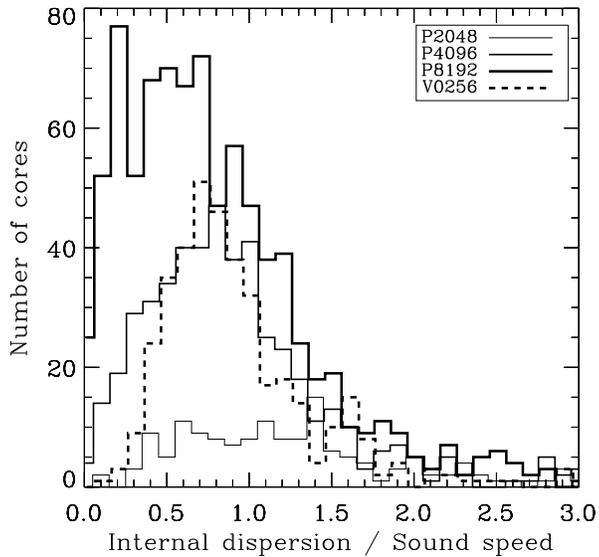,width=\columnwidth}
  \caption{Ratio of internal velocity dispersion and internal
    sound speed (i.e. the internal Mach number) for all cold clumps. 
    For all models, most of the clumps have subsonic internal motions.
    \label{f:dis_cs}}
\end{center}
\end{figure}

\begin{figure}
\begin{center}
  \epsfig{file=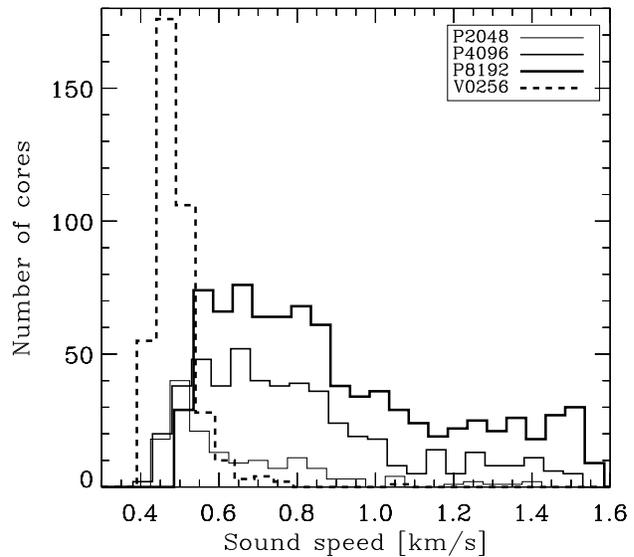,width=\columnwidth}
  \caption{Internal sound speed of the cold clumps ($T<300$K).
  The clumps found in model V0256 have systematically lower
  internal dispersions with a smaller spread. \label{f:soundhist}}
\end{center}
\end{figure}

\begin{figure}
\begin{center}
  \epsfig{file=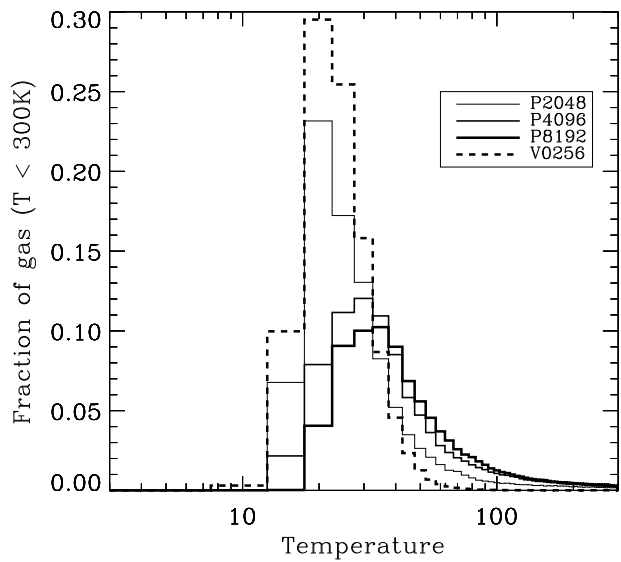,width=\columnwidth}
  \caption{Fractional mass as a function of temperature in the cold
  phase (not limited to clumps). There is relatively more low temperature ($T < 40$~K) gas in model V0256
  than in the Proteus models.
  \label{f:temphist}}
\end{center}
\end{figure}

With the sound speed in the cold gas $c_s\approx 0.75$~km~s$^{-1}$, 
the {\em internal} velocity dispersion of the cold clumps turn out
to be mostly subsonic (Fig.~\ref{f:dis_cs}). This confirms
earlier findings from numerical simulations \citep{2002ApJ...564L..97K,
2005A&A...433....1A,2006ApJ...648.1052H}, and should not come as a surprise
in view of the formation mechanism of the cold dense clumps. 
The Proteus models have a broader distribution 
of internal velocity dispersions 
than the VINE models,
and the peaks shift to smaller velocities with increasing resolution, indicating
that the thermal length scales are not completely resolved at the lowest
resolutions \citep[see also][]{2007A&A...465..431H}. The
distribution for the V0256 clumps is much narrower and peaks around $0.25$~km~s$^{-1}$.
For comparison we show the distribution of the internal sound speeds of
the individual clumps in Fig. \ref{f:soundhist}. Both distributions
peak around $0.45$ - $0.50$~km~s$^{-1}$. However, the clumps in the SPH
simulations tend to cluster at lower temperature ($T<100$~K) and there is less
gas at higher temperatures than in the grid simulations (Fig. \ref{f:temphist}). 
This is probably a result of SPH intrinsic over-cooling. 


\section{Summary and Conclusions} 
\label{SUMMARY}
Lagrangian hydrodynamical methods such as Smoothed Particle Hydrodynamics (SPH) 
are an attractive tool to simulate astrophysical problems with a wide range of spatial
scales, such as self-gravitating objects or thermally unstable flows. However,
due to the numerical implementations, the reliability of SPH for such applications
has been questioned. We compare a SPH implementation 
\citep[VINE, ][]{2009ApJS..184..326N,2009ApJS..184..298W}
and a grid-based method
\citep[Proteus, ][]{1993JCoPh.109...53P,1999A&AS..139..199S,2001JCoPh.171..289X,2004ApJ...603..165H}
in an application to flow-driven molecular cloud formation. 
This is a challenging astrophysical problem for any numerical method, because
of the turbulence, the high density and temperature contrasts, and the rapidly
shrinking scales of the cold clouds.

The morphologies between particle and grid based methods agree well globally, yet the cold gas
structures are more fragmented in the SPH models (Fig.~\ref{temp_dens_snap}).
Temperature and density gradients in SPH tend to be ``rounder''
than in the grid method as a consequence of deriving physical quantities
by averaging over a set of nearest neighbors. However, this drawback is
countered by the higher spatial resolution, yielding more filigree structure
in the SPH models.

The mass fractions of gas in different temperature regimes differ slightly (Fig.~\ref{time_vs_massfrac}).
SPH models start collecting mass in the cold stable regime later than the grid
models, leading to a constant offset in the mass with time. This is a consequence
of the strong overheating in the initial flow collision due to SPH's artificial
viscosity. The slopes of the ClMFs (Fig.~\ref{f:cmf}) are consistent with 
results from earlier numerical models \citep{2007A&A...465..445H} with similar
physics. They are also consistent with observational results 
\citep{2002A&A...384..225S,2007ApJ...655..351L,2009ApJ...698..324R},
yet such a comparison should not be over-interpreted because of the restricted
dimensionality of our models as well as the fact that self-gravity will have
affected the evolution of the cloud at the time we chose to analyze the models
\citep{2008ApJ...689..290H}.

The line-of-sight velocity distributions have similar widths (Fig.~\ref{f:losvd}), namely 
$\approx 6.8$~km~s$^{-1}$ for VINE and $\approx 8.5$~km~s$^{-1}$ for Proteus, rendering
the gas nominally supersonic, while both methods find that the {\em internal} velocity
dispersion of the cold clumps is subsonic \citep[see also][]{2002ApJ...564L..97K,
2005A&A...433....1A,2006ApJ...648.1052H}. The slightly
narrower distribution for SPH is a consequence of the higher
resolution, emphasizing the thermal instabilities over the dynamical ones.

The effects of overcooling \citep{1999ApJ...521L..99P} do not seem to affect the 
overall mass budget (Fig.~\ref{time_vs_massfrac}) within the cold stable regime, 
but they do manifest themselves in the low-temperature tail of that phase 
(Figs.~\ref{f:soundhist} and \ref{f:temphist}). This suggests that the TI controlling 
the transition from the warm stable to the cold stable regime is not strongly affected. 

Our results suggest that the SPH implementation in VINE is 
suitable to model thermally bistable, dynamical environments such as occurring
in the flow-driven formation of molecular clouds. Yet, quantitative predictions will slightly
differ between particle and grid methods, with a bias of the VINE results to lower temperatures 
and smaller fragments in the cold gas, but a lower total mass in the cold, thermally stable regime compared
to Proteus. In terms of mass spectra, an initial particle resolution of $256^2$ is (at later stages) equivalent
or better than a fixed grid of $4096^2$ cells, with promising consequences for the determination
of e.g. clump mass functions.

\section*{Acknowledgments}
We thank the anonymous referee for a very helpful report that led to a clearer
focus of the paper.
Simulations were run on the SGI-Altix at the University Observatory Munich,
partly funded by the DFG cluster of excellence ``Origin and Structure of the Universe'',
and on the local PC cluster Star of the 
Department of Astronomy at the University of Michigan,
maintained by J.~Hallum. FH acknowledges support the 
University of Michigan, by NASA grant NNG06GJ32G, and by NSF grant AST 0807305.
SW acknowledges the support of the Marie Curie RTN \textsc{Constellation} 
(MRTN-CT-2006-035890).

\bibliographystyle{mn2e}
\bibliography{./references.bib}

\label{lastpage}
\end{document}